# $\mathbb{SE}(3)$ based Extended Kalman Filter for Spacecraft Attitude Estimation

Lubin Chang[*]

*Naval University of Engineering, Wuhan, 430033, China*

**Abstract**: In this paper, the spacecraft attitude estimation problem has been investigated making use of the concept of matrix Lie group. Through formulation of the attitude and gyroscope bias as elements of $\mathbb{SE}(3)$, the corresponding extended Kalman filter, termed as $\mathbb{SE}(3)$-EKF, has been derived. It is shown that the resulting $\mathbb{SE}(3)$-EKF is just the newly-derived geometric extended Kalman filter (GEKF) for spacecraft attitude estimation. This provides a new perspective on the GEKF besides the common frame errors definition. Moreover, the $\mathbb{SE}(3)$-EKF with reference frame attitude error is also derived and the resulting algorithm bears much resemblance to the right invariant EKF.

## 1. Introduction

Spacecraft attitude estimation using vector observations has drawn much attention since the 1960s. For this problem, two facts have been widely approved, that is, gyroscopes are used in "dynamic model replacement mode" [1] and the utility of employing quaternions for attitude parameterization is preferred [2]. For the first fact, since the gyroscopes have been used, the gyroscope bias is always estimated coupled with the attitude. For the second fact, many filtering methods have been devoted to handling the constraints inherent to non-minimal attitude representation. The extended Kalman filter (EKF) in multiplicative form (MEKF) is the workhorse for on-board attitude estimation due to its flexibility and computational efficiency [2-4]. However, the EKF may prone to divergence when dynamics or measurement models are highly nonlinear, or there is no good a priori state estimate. In this case, some advanced nonlinear filtering algorithms can be applied [5-10]. For more details about the advanced attitude estimators until 2006, the comprehensive and excellent survey paper [4] can be referred. It should be noted that most of these previous attitude estimators put much effort on the attitude part of the state vector. Non-attitude state vectors have still been handled in the general vector-space. From the perspective of matrix Lie group, these attitude estimators can be viewed as the extensions of general filters on *special orthogonal group*, i.e. $\mathbb{SO}(3)$.

---

[*] Associated Professor, Department of Navigation Engineering, No. 717 Jiefang Road, Wuhan, 430033, China. changlubin@163.com

In recent few years, a filter called geometric extended Kalman filter (GEKF) has been well studied for attitude estimation [11-14]. It has been demonstrated that the GEKF can outperform the MEKF. The main concept of GEKF is a new state-error definition with involved errors expressed in a common frame. The common frame is established making use of the estimated attitude error in each filtering recursion. Compared with the traditional state-error in vector-space, the frame consistent errors are more representative of the real errors experienced by the system. The frame consistent error is nonlinear due to error-attitude coupling with the states. In contrast with the linear error in vectorspace, such nonlinear error can be viewed as on manifold [15-18]. The above viewpoint motives us to revisit the GEKF making use of the concept of Lie group whose operations are just performed on manifold. In this note, we formulate the attitude matrix and gyroscope bias as elements of *special Euclidean group* $\mathbb{SE}(3)$, a representative of the Lie group. The detailed filtering equations are derived making use of matrix Lie group and Lie algebra theories for spacecraft attitude estimation. The resulting attitude estimation algorithm is termed as $\mathbb{SE}(3)-\text{EKF}$. It is revealed that the GEKF with body frame attitude error bears much resemblance to, and thus can be viewed as, a kind of the $\mathbb{SE}(3)-\text{EKF}$. This provides another perspective on the GEKF besides the common frame errors definition. Moreover, the $\mathbb{SE}(3)-\text{EKF}$ with reference frame attitude error is also derived and the resulting algorithm bears much resemblance to the right invariant EKF (RIEKF) proposed in [19].

The organization of this paper proceeds as follows. In Section 2, the matrix Lie group and spacecraft attitude estimation model using vector observations have been introduced as mathematical preliminaries. In Section 3, the $\mathbb{SE}(3)-\text{EKF}$ with body frame attitude error is derived. Its relationship with GEKF is also revealed and discussed. The $\mathbb{SE}(3)-\text{EKF}$ with reference frame attitude error is derived in Section 3. Some conclusions are drawn in the final section.

## 2. Mathematical Preliminaries

In this section, the mathematical preliminaries related with matrix Lie group, Lie algebra and spacecraft attitude estimation model are provided, which will be used to derive the $\mathbb{SE}(3)-\text{EKF}$ in the following sections.

**2.1 Matrix Lie Groups and Lie Algebra**

The *special Euclidean group* $\mathbb{SE}(3)$, which is always used to represent the poses, is simply the set of valid transformation matrices as [20]

$$\mathbb{SE}(3) = \{\Psi(\mathbf{R},\mathbf{t}); \mathbf{R}\in\mathbb{SO}(3), \mathbf{t}\in\mathbb{R}^3\} \tag{1}$$

where $\mathbb{SO}(3)$ is *special orthogonal group*, representing rotations. $\mathbb{SE}(3)$ and $\mathbb{SO}(3)$

are both types of matrix Lie group. The map $\Psi$ is given by

$$\Psi:(\mathbf{R},\mathbf{t}) \mapsto \underbrace{\begin{bmatrix} \mathbf{R} & \mathbf{t} \\ \mathbf{0}_{1\times 3} & 1 \end{bmatrix}}_{\mathbf{T}} \tag{2}$$

The inverse of an element of $\mathbb{SE}(3)$ is given by

$$\mathbf{T}^{-1} = \begin{bmatrix} \mathbf{R} & \mathbf{t} \\ \mathbf{0}_{1\times 3} & 1 \end{bmatrix}^{-1} = \begin{bmatrix} \mathbf{R}^T & -\mathbf{R}^T\mathbf{t} \\ \mathbf{0}_{1\times 3} & 1 \end{bmatrix} \in \mathbb{SE}(3) \tag{3}$$

With every matrix Lie group is associated a Lie algebra. The Lie algebra associated with $\mathbb{SE}(3)$ is given by

$$\mathfrak{se}(3) = \left\{ \pounds_{\mathfrak{se}(3)}(\zeta) \in \mathbb{R}^{4\times 4} \,\middle|\, \zeta \in \mathbb{R}^6 \right\} \tag{4}$$

where

$$\pounds_{\mathfrak{se}(3)}(\zeta) = \pounds_{\mathfrak{se}(3)}\begin{pmatrix} \boldsymbol{\varphi} \\ \boldsymbol{\mu} \end{pmatrix} = \begin{pmatrix} \boldsymbol{\varphi} \\ \boldsymbol{\mu} \end{pmatrix}^{\wedge} = \begin{bmatrix} (\boldsymbol{\varphi}\times) & \boldsymbol{\mu} \\ \mathbf{0}_{1,3} & 0 \end{bmatrix} \tag{5}$$

The relation between a matrix Lie group to its associated Lie algebra is given by the *exponential map*. The elements of $\mathbb{SE}(3)$ are related to the elements of $\mathfrak{se}(3)$ through

$$\mathbf{T} = \exp(\zeta^{\wedge}) = \sum_{n=0}^{\infty} \frac{1}{n!}(\zeta^{\wedge})^n = \begin{bmatrix} \mathbf{C} & \mathbf{J}\boldsymbol{\mu} \\ \mathbf{0}_{1\times 3} & 1 \end{bmatrix} \in \mathbb{SE}(3) \tag{6}$$

where

$$\mathbf{C} = \exp(\boldsymbol{\varphi}\times) = \sum_{n=0}^{\infty} \frac{1}{n!}(\boldsymbol{\varphi}\times)^n \tag{7}$$

$$\mathbf{J} = \sum_{n=0}^{\infty} \frac{1}{(n+1)!}(\boldsymbol{\varphi}\times)^n \tag{8}$$

Let $\boldsymbol{\varphi} = \varphi\mathbf{a}$ where $\varphi$ is the angle of rotation and $\mathbf{a} = \boldsymbol{\varphi}/\varphi$ is the unit-length axis of rotation. A direct series expression for $\mathbf{T}$ from the exponential map can also be obtained as

$$\exp(\zeta^\wedge) = \mathbf{I}_{4\times 4} + \zeta^\wedge + \left(\frac{1-\cos\varphi}{\varphi^2}\right)(\zeta^\wedge)^2 + \left(\frac{\varphi-\sin\varphi}{\varphi^3}\right)(\zeta^\wedge)^3 \tag{9}$$

**2.2 Spacecraft Attitude Estimation Model**

Denote the attitude quaternion as $\mathbf{q} = \begin{bmatrix} \boldsymbol{\rho}^T & q_4 \end{bmatrix}^T$ with $\boldsymbol{\rho} = \begin{bmatrix} q_1 & q_2 & q_3 \end{bmatrix}^T$ being the vector component. The spacecraft attitude kinematics model in terms of quaternion is given by

$$\dot{\mathbf{q}}(t) = \frac{1}{2}\Xi[\mathbf{q}(t)]\boldsymbol{\omega}(t) \tag{10}$$

where

$$\Xi[\mathbf{q}] = \begin{bmatrix} q_4 \mathbf{I}_{3\times 3} + [\boldsymbol{\rho}\times] \\ -\boldsymbol{\rho}^T \end{bmatrix} \tag{11}$$

$\boldsymbol{\omega}$ is the angular rate measured by the gyroscopes and governed by

$$\begin{aligned} \tilde{\boldsymbol{\omega}} &= \boldsymbol{\omega} + \boldsymbol{\beta} + \boldsymbol{\eta}_v \\ \dot{\boldsymbol{\beta}} &= \boldsymbol{\eta}_u \end{aligned} \tag{12}$$

where $\boldsymbol{\beta}$ is the constant gyroscope bias. $\boldsymbol{\eta}_v$ and $\boldsymbol{\eta}_u$ are assumed to be zero-mean, Gaussian white-noise processes.

The measurement model with $n$ vector observations is given by

$$\mathbf{y}_k = \begin{bmatrix} \mathbf{A}(\mathbf{q})\mathbf{r}_1 \\ \vdots \\ \mathbf{A}(\mathbf{q})\mathbf{r}_n \end{bmatrix}_{t_k} + \begin{bmatrix} \boldsymbol{v}_1 \\ \vdots \\ \boldsymbol{v}_n \end{bmatrix}_{t_k} \tag{13}$$

where $\mathbf{A}(\mathbf{q})$ is the attitude matrix corresponding to quaternion $\mathbf{q}$. $\boldsymbol{v}_i$ is the measurement white noise for the $i$-th vector observation.

### 3. $\mathbb{SE}(3)$–EKF with Body Frame Attitude Error

In this note, we follow the attitude error terminology in [3, 11] with emphases on the expressed frames of the attitude error. More specifically, given true quaternion related attitude matrix $\mathbf{A}(\mathbf{q})$ and its estimate $\mathbf{A}(\hat{\mathbf{q}})$, the body frame attitude error is given by $\mathbf{A}(\mathbf{q})\mathbf{A}(\hat{\mathbf{q}})^{-1}$ and the reference attitude error is given by $\mathbf{A}(\hat{\mathbf{q}})^{-1}\mathbf{A}(\mathbf{q})$. This is a little difference with the terminology used in Lie group where $\mathbf{A}(\mathbf{q})\mathbf{A}(\hat{\mathbf{q}})^{-1}$ is termed as *right* error and $\mathbf{A}(\hat{\mathbf{q}})^{-1}\mathbf{A}(\mathbf{q})$ as *left* error. Due to different representations of the attitude quaternion, the *right* attitude error may be expressed in reference frame [21]. With the attitude error terminology in [3, 11], we will firstly derive the $\mathbb{SE}(3)-\text{EKF}$ with body frame attitude error definition in this section.

Define the state as $\mathbf{X}=[\mathbf{A}(\mathbf{q}),\boldsymbol{\beta}]$. The transformation matrix corresponding to the state is given by

$$\chi = \Psi(\mathbf{X}) = \begin{bmatrix} \mathbf{A}(\mathbf{q}) & \boldsymbol{\beta} \\ \mathbf{0}_{1\times 3} & 1 \end{bmatrix} \tag{14}$$

Assume the estimate of $\chi$ is $\hat{\chi}$. The estimation error is defined as

$$\gamma = \chi\hat{\chi}^{-1} = \begin{bmatrix} \mathbf{A}(\mathbf{q})\mathbf{A}(\hat{\mathbf{q}})^{-1} & \boldsymbol{\beta} - \mathbf{A}(\mathbf{q})\mathbf{A}(\hat{\mathbf{q}})^{-1}\hat{\boldsymbol{\beta}} \\ \mathbf{0}_{1,3} & 1 \end{bmatrix} \tag{15}$$

In the derivation of (15), the equation (3) has been used. It is clearly shown that the estimation error is also *well-defined* on the group $\mathbb{SE}(3)$. According to relationship between matrix Lie group to its associated Lie algebra, $\gamma$ can also be obtained as

$$\gamma = \exp(\mathbf{dx}^\wedge) \tag{16}$$

where $\mathbf{dx} \in \mathbb{R}^6$. The linearized error dynamic model of $\mathbf{dx}$ can be given by

$$\mathbf{d\dot{x}}(t) = \mathbf{F}(\hat{\chi}(t),t)\mathbf{dx}(t) + \mathbf{G}(\hat{\chi}(t),t)\mathbf{w}(t) \tag{17}$$

where $\mathbf{F}(\hat{\chi}(t),t)$ and $\mathbf{G}(\hat{\chi}(t),t)$ are the Jacobians with respect to the estimation error and system noise on matrix Lie group, respectively. Once the associated Jacobians are calculated, the general EKF framework can be applied to update the error-state.

According to (9), a first order approximation of $\gamma$ is given by

$$\gamma = \exp\left(\mathbf{dx}^{\wedge}\right) \approx \mathbf{I}_{4\times 4} + \mathbf{dx}^{\wedge} \tag{18}$$

According to [13], the first-order approximation of the error-attitude matrix is given by

$$\mathbf{A}(\delta\mathbf{q}) = \mathbf{A}(\mathbf{q})\mathbf{A}(\hat{\mathbf{q}})^{-1} \approx \mathbf{I}_{3\times 3} - (\delta\boldsymbol{\alpha}\times) \tag{19}$$

where $\delta\boldsymbol{\alpha}$ has components of roll, pitch and yaw error angles for any rotation sequence.

Substituting (19) into (15) yields

$$\gamma \approx \begin{bmatrix} \mathbf{I}_{3\times 3} - (\delta\boldsymbol{\alpha}\times) & \boldsymbol{\beta} - \hat{\boldsymbol{\beta}} + \delta\boldsymbol{\alpha}\times\hat{\boldsymbol{\beta}} \\ \mathbf{0}_{1\times 3} & 1 \end{bmatrix} \tag{20}$$

Comparing (18) and (20), the definite expression of $\mathbf{dx}$ can be obtained as

$$\mathbf{dx} = \begin{bmatrix} -\delta\boldsymbol{\alpha} \\ \boldsymbol{\beta} - \hat{\boldsymbol{\beta}} + \delta\boldsymbol{\alpha}\times\hat{\boldsymbol{\beta}} \end{bmatrix} = \begin{bmatrix} -\delta\boldsymbol{\alpha} \\ \Delta\boldsymbol{\beta} + \delta\boldsymbol{\alpha}\times\hat{\boldsymbol{\beta}} \end{bmatrix} = \begin{bmatrix} \mathbf{d}\boldsymbol{\alpha} \\ \mathbf{d}\boldsymbol{\beta} \end{bmatrix} \tag{21}$$

where $\mathbf{d}\boldsymbol{\alpha}$ and $\mathbf{d}\boldsymbol{\beta}$ is clearly defined in (21).

According to [3], the explicit models of $\delta\boldsymbol{\alpha}$ and $\Delta\boldsymbol{\beta}$ are given by

$$\delta\dot{\boldsymbol{\alpha}} = -(\hat{\boldsymbol{\omega}}\times)\delta\boldsymbol{\alpha} - (\Delta\boldsymbol{\beta} + \boldsymbol{\eta}_v) \tag{22}$$

$$\Delta\dot{\boldsymbol{\beta}} = \boldsymbol{\eta}_u \tag{23}$$

where $\hat{\boldsymbol{\omega}} = \tilde{\boldsymbol{\omega}} - \hat{\boldsymbol{\beta}}$. Next, we will derive the models of $\mathbf{d}\boldsymbol{\alpha}$ and $\mathbf{d}\boldsymbol{\beta}$ according to their definitions in (21) using the existing models (22) and (23).

Rewrite the definition of $\mathbf{d}\boldsymbol{\beta}$ as

$$\mathbf{d}\boldsymbol{\beta} = \Delta\boldsymbol{\beta} + \delta\boldsymbol{\alpha}\times\hat{\boldsymbol{\beta}} \tag{24}$$

Taking the time derivative of (24) gives

$$\begin{aligned}
\mathbf{d}\dot{\boldsymbol{\beta}} &= \Delta\dot{\boldsymbol{\beta}} + \delta\dot{\boldsymbol{\alpha}} \times \hat{\boldsymbol{\beta}} + \delta\boldsymbol{\alpha} \times \dot{\hat{\boldsymbol{\beta}}} \\
&= \boldsymbol{\eta}_u + \left(-(\hat{\boldsymbol{\omega}}\times)\delta\boldsymbol{\alpha} - (\Delta\boldsymbol{\beta} + \boldsymbol{\eta}_v)\right) \times \hat{\boldsymbol{\beta}} \\
&= (\hat{\boldsymbol{\beta}}\times)(\hat{\boldsymbol{\omega}}\times)\delta\boldsymbol{\alpha} + (\hat{\boldsymbol{\beta}}\times)\Delta\boldsymbol{\beta} + (\hat{\boldsymbol{\beta}}\times)\boldsymbol{\eta}_v + \boldsymbol{\eta}_u \\
&= (\hat{\boldsymbol{\beta}}\times)(\hat{\boldsymbol{\omega}}\times)\delta\boldsymbol{\alpha} + (\hat{\boldsymbol{\beta}}\times)(\mathbf{d}\boldsymbol{\beta} - \delta\boldsymbol{\alpha}\times\hat{\boldsymbol{\beta}}) + (\hat{\boldsymbol{\beta}}\times)\boldsymbol{\eta}_v + \boldsymbol{\eta}_u \\
&= -(\hat{\boldsymbol{\beta}}\times)(\hat{\boldsymbol{\omega}}\times)\mathbf{d}\boldsymbol{\alpha} + (\hat{\boldsymbol{\beta}}\times)\mathbf{d}\boldsymbol{\beta} + (\hat{\boldsymbol{\beta}}\times)\boldsymbol{\eta}_v + \boldsymbol{\eta}_u
\end{aligned} \quad (25)$$

In (25), we have made use of the definition $\mathbf{d}\boldsymbol{\alpha} = -\delta\boldsymbol{\alpha}$.

Substituting (24) into (22) gives

$$\begin{aligned}
\delta\dot{\boldsymbol{\alpha}} &= -(\hat{\boldsymbol{\omega}}\times)\delta\boldsymbol{\alpha} - \left(\mathbf{d}\boldsymbol{\beta} - \delta\boldsymbol{\alpha}\times\hat{\boldsymbol{\beta}} + \boldsymbol{\eta}_v\right) \\
&= -\left((\hat{\boldsymbol{\omega}} + \hat{\boldsymbol{\beta}})\times\right)\delta\boldsymbol{\alpha} - \mathbf{d}\boldsymbol{\beta} - \boldsymbol{\eta}_v \\
&= -(\tilde{\boldsymbol{\omega}}\times)\delta\boldsymbol{\alpha} - \mathbf{d}\boldsymbol{\beta} - \boldsymbol{\eta}_v
\end{aligned} \quad (26)$$

According to the definition $\mathbf{d}\boldsymbol{\alpha} = -\delta\boldsymbol{\alpha}$, the model of the definition $\mathbf{d}\boldsymbol{\alpha}$ can be readily obtained as

$$\mathbf{d}\dot{\boldsymbol{\alpha}} = -(\tilde{\boldsymbol{\omega}}\times)\mathbf{d}\boldsymbol{\alpha} + \mathbf{d}\boldsymbol{\beta} + \boldsymbol{\eta}_v \quad (27)$$

Based on (25) and (27), we can easily derive the Jacobians as

$$\mathbf{F}(\hat{\boldsymbol{\chi}}(t),t) = \begin{bmatrix} -(\tilde{\boldsymbol{\omega}}\times) & \mathbf{I}_{3\times3} \\ -(\hat{\boldsymbol{\beta}}\times)(\hat{\boldsymbol{\omega}}\times) & (\hat{\boldsymbol{\beta}}\times) \end{bmatrix} \quad (28)$$

$$\mathbf{G}(\hat{\boldsymbol{\chi}}(t),t) = \begin{bmatrix} \mathbf{I}_{3\times3} & \mathbf{0}_{3\times3} \\ (\hat{\boldsymbol{\beta}}\times) & \mathbf{I}_{3\times3} \end{bmatrix} \quad (29)$$

Next, we will derive the measurement transition matrix for the error-state $\mathbf{dx}$. For a single sensor the true and estimated body vectors are given by

$$\mathbf{b} = \mathbf{A}(\mathbf{q})\mathbf{r} \quad (30)$$

$$\hat{\mathbf{b}}^- = \mathbf{A}(\hat{\mathbf{q}}^-)\mathbf{r} \quad (31)$$

Subtracting (31) from (30) and making use of the approximation in (19) yield

$$\begin{aligned}
\Delta \mathbf{b} &= \mathbf{A}(\delta \mathbf{q})\mathbf{A}(\hat{\mathbf{q}}^-)\mathbf{r} - \mathbf{A}(\hat{\mathbf{q}}^-)\mathbf{r} \\
&= (\mathbf{I}_{3\times 3} - (\delta \boldsymbol{\alpha} \times))\mathbf{A}(\hat{\mathbf{q}}^-)\mathbf{r} - \mathbf{A}(\hat{\mathbf{q}}^-)\mathbf{r} \\
&= \left[\mathbf{A}(\hat{\mathbf{q}}^-)\mathbf{r} \times\right]\delta \boldsymbol{\alpha} = -\left[\mathbf{A}(\hat{\mathbf{q}}^-)\mathbf{r} \times\right]d\boldsymbol{\alpha}
\end{aligned} \tag{31}$$

Then, the linearized measurement model for multiple vector observations is given by

$$\mathbf{y} = \mathbf{H}d\mathbf{x} = \begin{bmatrix} -\left[\mathbf{A}(\hat{\mathbf{q}}^-)\mathbf{r}_1 \times\right] & \mathbf{0}_{3\times 3} \\ \vdots & \vdots \\ -\left[\mathbf{A}(\hat{\mathbf{q}}^-)\mathbf{r}_n \times\right] & \mathbf{0}_{3\times 3} \end{bmatrix} d\mathbf{x} \tag{32}$$

Given the above derived state-space model, the $\mathbb{SE}(3)$–EKF for spacecraft attitude estimation using vector observations is summarized in **Algorithm 1**.

---

**Algorithm 1:** $\mathbb{SE}(3)$–EKF with Body Frame Attitude Error

---
Initialize:
$\hat{\mathbf{X}}(t_0) = \left[\mathbf{A}(\hat{\mathbf{q}}(t_0)), \hat{\boldsymbol{\beta}}(t_0)\right] = \left[\mathbf{A}(\hat{\mathbf{q}}_0), \hat{\boldsymbol{\beta}}_0\right]$
$\mathbf{P}(t_0) = E\left(d\mathbf{x}_0 d\mathbf{x}_0^T\right)$
Gain:
$\mathbf{K}_k = \mathbf{P}_{k|k-1}\mathbf{H}_k^T \left[\mathbf{H}_k \mathbf{P}_{k|k-1}\mathbf{H}_k^T + \mathbf{R}_k\right]^{-1}$
Update:
$d\hat{\mathbf{x}}_k = \mathbf{K}_k \left(\mathbf{y}_k - h(\hat{\mathbf{X}}_{k|k-1})\right)$
$\hat{\boldsymbol{\chi}}_k = \exp(d\hat{\mathbf{x}}_k^{\wedge})\Psi\left(\hat{\mathbf{X}}_{k|k-1}\right)$
$\hat{\mathbf{X}}_k = \Psi^{-1}(\hat{\boldsymbol{\chi}}_k)$
Propagation:
$\hat{\boldsymbol{\omega}}(t) = \tilde{\boldsymbol{\omega}}(t) - \hat{\boldsymbol{\beta}}(t)$
$\dot{\hat{\mathbf{q}}}(t) = \frac{1}{2}\Xi(\hat{\mathbf{q}}(t))\hat{\boldsymbol{\omega}}(t)$
$\hat{\boldsymbol{\chi}}(t) = \Psi(\hat{\mathbf{X}}(t)) = \begin{bmatrix} \mathbf{A}(\hat{\mathbf{q}}(t)) & \hat{\boldsymbol{\beta}}(t) \\ \mathbf{0}_{1\times 3} & 1 \end{bmatrix}$
$\dot{\mathbf{P}}(t) = \mathbf{F}(\hat{\boldsymbol{\chi}}(t),t)\mathbf{P}(t) + \mathbf{P}(t)\mathbf{F}^T(\hat{\boldsymbol{\chi}}(t),t)$
$\qquad + \mathbf{G}(\hat{\boldsymbol{\chi}}(t),t)\mathbf{Q}_{k-1}\mathbf{G}(\hat{\boldsymbol{\chi}}(t),t)$

---

In Algorithm 1, $\mathbf{H}_k$ and $h(\hat{\mathbf{X}}_{k|k-1})$ are given by

$$\mathbf{H}_k = \begin{bmatrix} -\left[\mathbf{A}(\hat{\mathbf{q}}_{k|k-1})\mathbf{r}_{k1} \times\right] & \mathbf{0}_{3\times 3} \\ \vdots & \vdots \\ -\left[\mathbf{A}(\hat{\mathbf{q}}_{k|k-1})\mathbf{r}_{kn} \times\right] & \mathbf{0}_{3\times 3} \end{bmatrix} \tag{33}$$

$$h\left(\hat{\mathbf{X}}_{k|k-1}\right) = \begin{bmatrix} \mathbf{A}\left(\hat{\mathbf{q}}_{k|k-1}\right)\mathbf{r}_{k1} \\ \vdots \\ \mathbf{A}\left(\hat{\mathbf{q}}_{k|k-1}\right)\mathbf{r}_{kn} \end{bmatrix} \qquad (34)$$

The transformation matrix estimate

$$\hat{\boldsymbol{\chi}}_k = \exp\left(\mathbf{d}\hat{\mathbf{x}}_k^{\wedge}\right)\Psi\left(\hat{\mathbf{X}}_{k|k-1}\right) \qquad (35)$$

is corresponding to the transformation matrix error definition in (15). If we divide the error-state $\mathbf{dx}$ as $\mathbf{d}\hat{\mathbf{x}}_k = \begin{bmatrix} \mathbf{d}\hat{\boldsymbol{\alpha}}_k^T & \mathbf{d}\hat{\boldsymbol{\beta}}_k^T \end{bmatrix}^T$, according to the first order approximation of exponential map in (18), (35) can be further approximated as

$$\begin{aligned}
\hat{\boldsymbol{\chi}}_k &\approx \left(\mathbf{I}_{4\times 4} + \mathbf{d}\hat{\mathbf{x}}_k^{\wedge}\right)\Psi\left(\hat{\mathbf{X}}_{k|k-1}\right) \\
&= \begin{bmatrix} \left[\mathbf{I}_{3\times 3} + \left(\mathbf{d}\hat{\boldsymbol{\alpha}}_k \times\right)\right]\mathbf{A}\left(\hat{\mathbf{q}}_{k|k-1}\right) & \left[\mathbf{I}_{3\times 3} + \left(\mathbf{d}\hat{\boldsymbol{\alpha}}_k \times\right)\right]\hat{\boldsymbol{\beta}}_{k|k-1} + \mathbf{d}\hat{\boldsymbol{\beta}}_k \\ \mathbf{0}_{1\times 3} & 1 \end{bmatrix}
\end{aligned} \qquad (36)$$

If we let

$$\mathbf{A}\left(\delta\hat{\mathbf{q}}_k\right) = \left[\mathbf{I}_{3\times 3} + \left(\mathbf{d}\hat{\boldsymbol{\alpha}}_k \times\right)\right] = \left[\mathbf{I}_{3\times 3} - \left(\delta\hat{\boldsymbol{\alpha}}_k \times\right)\right] \qquad (37)$$

The estimate Lie matrix $\hat{\mathbf{X}}_k$ is given by

$$\hat{\mathbf{X}}_k = \left[\mathbf{A}\left(\delta\hat{\mathbf{q}}_k\right)\mathbf{A}\left(\hat{\mathbf{q}}_{k|k-1}\right), \mathbf{A}\left(\delta\hat{\mathbf{q}}_k\right)\hat{\boldsymbol{\beta}}_{k|k-1} + \mathbf{d}\hat{\boldsymbol{\beta}}_k\right] \qquad (38)$$

According to the aforementioned filtering recursion, if we use $\boldsymbol{\delta\alpha}$ instead of $\mathbf{d\alpha}$ to represent the attitude error, the resulting state-space model is identical with that of the GEKF. That is to say, the derived $\mathbb{SE}(3)$-EKF with body frame attitude error is just a minor variant of the multiplicative form of the GEKF (set aside the covariance update for the global state update). In other word, the GEKF for attitude estimation can be viewed as a kind of EKF on group $\mathbb{SE}(3)$. This recognition provides another viewpoint on the geometric error-state used in GEKF. The common frame interpretation is virtually the physical meaning of the geometric error-state and the

$\mathbb{SE}(3)$ based derivation can provide a mathematical perspective.

With the above perspective, the traditional MEKF can also be viewed as a kind of $\mathbb{SO}(3)-\text{EKF}$. According to the invariance theory [15-18], $\mathbb{SE}(3)-\text{EKF}$ bears better invariance properties than $\mathbb{SO}(3)-\text{EKF}$ and therefore, can always outperforms the $\mathbb{SO}(3)-\text{EKF}$. This conclusion has been well demonstrated through the comparisons between traditional MEKF and GEKF in [11]. The main utility of the $\mathbb{SE}(3)$ perspective lies in the nonlinear state-error construction. When other parameters, such as gyroscope scale factors and misalignments, are added to the state vector, we can construct the corresponding nonlinear state-error according to the procedures (14) and (15). Actually, in practical applications, there may be no need to define all the involved parameters errors on manifold and some can be defined still in vector-space [22-24]. In this respect, we should firstly determine the group structure with the form $\mathbb{SE}_n(3)\times\mathbb{R}^{3\times m}$ where $n$ is the number of the three-dimensional parameters whose errors should be defined on manifold while $m$ is the number of those should be defined in vector-space. With the defined group structure, we can therefore determine the error-state and the corresponding state-space model.

It should be noted that there is no need to use (35) to update the state and the first-order approximation in (38) is adequate. This is because that when deriving the linear state-space model, first-order approximations (18) and (19) have been used. Rigorously updating state using (35) isn't likely to add any additional accuracy.

## 4. $\mathbb{SE}(3)-\text{EKF}$ with Reference Frame Attitude Error

Instead of definition of the attitude error in body frame, it can also be defined in reference frame [25]. In contrast with (15), an alternative form of transformation matrix error is defined as

$$\gamma = \hat{\chi}^{-1}\chi = \begin{bmatrix} \mathbf{A}(\hat{\mathbf{q}})^{-1}\mathbf{A}(\mathbf{q}) & \mathbf{A}(\hat{\mathbf{q}})^{-1}(\boldsymbol{\beta}-\hat{\boldsymbol{\beta}}) \\ \mathbf{0}_{1,3} & 1 \end{bmatrix} \tag{39}$$

Making use of the approximation in (19), (39) can be approximated as

$$\gamma \approx \begin{bmatrix} \mathbf{I}_{3\times 3} - (\boldsymbol{\delta\alpha}\times) & \mathbf{A}(\hat{\mathbf{q}})^{-1}(\boldsymbol{\beta}-\hat{\boldsymbol{\beta}}) \\ \mathbf{0}_{1\times 3} & 1 \end{bmatrix} \tag{40}$$

Since $\gamma \in \mathbb{SE}(3)$, it can also be obtained as

$$\gamma = \exp(\mathbf{dx}^{\wedge}) \tag{41}$$

where $\mathbf{dx} \in \mathbb{R}^6$. Making use of the first-order approximation in (18) and comparing the approximated result of (41) with (40), we can obtain

$$\mathbf{dx} = \begin{bmatrix} -\boldsymbol{\delta\alpha} \\ \mathbf{A}(\hat{\mathbf{q}})^{-1}(\boldsymbol{\beta}-\hat{\boldsymbol{\beta}}) \end{bmatrix} = \begin{bmatrix} -\boldsymbol{\delta\alpha} \\ \mathbf{A}(\hat{\mathbf{q}})^{-1}\Delta\boldsymbol{\beta} \end{bmatrix} = \begin{bmatrix} \mathbf{d\alpha} \\ \mathbf{d\beta} \end{bmatrix} \tag{42}$$

where $\mathbf{d\alpha}$ and $\mathbf{d\beta}$ is clearly defined in (42).

Next we will derive the state-space model of $\mathbf{dx}$ making use of the similar procedures in Section III. The state-space model with reference frame attitude error for the traditional MEKF is derived in Appendix. Based on the definitions of $\mathbf{d\alpha}$ and $\mathbf{d\beta}$ in (42) and the attitude error model (A10), we can easily obtain the model of $\mathbf{d\alpha}$ as

$$\mathbf{d\dot{\alpha}} = \mathbf{d\beta} + \mathbf{A}(\hat{\mathbf{q}})^T \boldsymbol{\eta}_v \tag{43}$$

The model of $\mathbf{d\beta}$ is given by

$$\begin{aligned}
\mathbf{d\dot{\beta}} &= \dot{\mathbf{A}}(\hat{\mathbf{q}})^T \Delta\boldsymbol{\beta} + \mathbf{A}(\hat{\mathbf{q}})^T \Delta\dot{\boldsymbol{\beta}} \\
&= \mathbf{A}(\hat{\mathbf{q}})^T (\hat{\boldsymbol{\omega}}\times)\Delta\boldsymbol{\beta} + \mathbf{A}(\hat{\mathbf{q}})^T \boldsymbol{\eta}_u \\
&= \mathbf{A}(\hat{\mathbf{q}})^T (\hat{\boldsymbol{\omega}}\times)\mathbf{A}(\hat{\mathbf{q}})\mathbf{A}(\hat{\mathbf{q}})^T \Delta\boldsymbol{\beta} + \mathbf{A}(\hat{\mathbf{q}})^T \boldsymbol{\eta}_u \\
&= \left[\left(\mathbf{A}(\hat{\mathbf{q}})^T \hat{\boldsymbol{\omega}}\right)\times\right]\mathbf{d\beta} + \mathbf{A}(\hat{\mathbf{q}})^T \boldsymbol{\eta}_u
\end{aligned} \tag{44}$$

Based on (43) and (44), the state-space process model for $\mathbf{dx}$ is given by

$$\mathbf{d\dot{x}}(t) = \mathbf{F}(\hat{\chi}(t),t)\mathbf{dx}(t) + \mathbf{G}(\hat{\chi}(t),t)\mathbf{w}(t) \tag{45}$$

where

$$\mathbf{F}(\hat{\mathbf{x}}(t),t) = \begin{bmatrix} \mathbf{0}_{3\times 3} & \mathbf{I}_{3\times 3} \\ \mathbf{0}_{3\times 3} & \left[\left(\mathbf{A}(\hat{\mathbf{q}})^T \hat{\boldsymbol{\omega}}\right)\times\right] \end{bmatrix} \quad (46)$$

$$\mathbf{G}(t) = \begin{bmatrix} \mathbf{A}(\hat{\mathbf{q}})^T & \mathbf{0}_{3\times 3} \\ \mathbf{0}_{3\times 3} & \mathbf{A}(\hat{\mathbf{q}})^T \end{bmatrix} \quad (47)$$

According to (A19), the linearized measurement model for multiple vector observations is given by

$$\mathbf{y} = \mathbf{H}\mathbf{dx} = \begin{bmatrix} -\left[\left(\mathbf{A}(\hat{\mathbf{q}}^-)\mathbf{r}\right)\times\right]\mathbf{A}(\hat{\mathbf{q}}^-) & \mathbf{0}_{3\times 3} \\ \vdots & \vdots \\ -\left[\left(\mathbf{A}(\hat{\mathbf{q}}^-)\mathbf{r}\right)\times\right]\mathbf{A}(\hat{\mathbf{q}}^-) & \mathbf{0}_{3\times 3} \end{bmatrix} \mathbf{dx} \quad (48)$$

Given the above derived state-space model, the $\mathbb{SE}(3)$–EKF with reference frame attitude error can be readily presented. Since this algorithm possesses similar procedure with Algorithm 1, it has not been presented here for brevity. Besides the involved state-space model, the only difference lies in the transformation matrix estimate equation. Specially, it is updated according to

$$\hat{\boldsymbol{\chi}}_k = \Psi\left(\hat{\mathbf{X}}_{k|k-1}\right)\exp\left(\mathbf{d}\hat{\mathbf{x}}_k^{\wedge}\right) \quad (49)$$

Denote the error-state $\mathbf{dx}$ as $\mathbf{d}\hat{\mathbf{x}}_k = \begin{bmatrix} \mathbf{d}\hat{\boldsymbol{\alpha}}_k^T & \mathbf{d}\hat{\boldsymbol{\beta}}_k^T \end{bmatrix}^T$. Making use of the first order approximation of exponential map in (18), (49) can be further approximated as

$$\begin{aligned}\hat{\boldsymbol{\chi}}_k &\approx \Psi\left(\hat{\mathbf{X}}_{k|k-1}\right)\left(\mathbf{I}_{4\times 4} + \mathbf{d}\hat{\mathbf{x}}_k^{\wedge}\right) \\ &= \begin{bmatrix} \mathbf{A}\left(\hat{\mathbf{q}}_{k|k-1}\right)\left[\mathbf{I}_{3\times 3} + \left(\mathbf{d}\hat{\boldsymbol{\alpha}}_k\times\right)\right] & \hat{\boldsymbol{\beta}}_{k|k-1} + \mathbf{A}\left(\hat{\mathbf{q}}_{k|k-1}\right)\mathbf{d}\hat{\boldsymbol{\beta}}_k \\ \mathbf{0}_{1\times 3} & 1 \end{bmatrix}\end{aligned} \quad (50)$$

If we let

$$\mathbf{A}\left(\delta\hat{\mathbf{q}}_k\right) = \left[\mathbf{I}_{3\times 3} + \left(\mathbf{d}\hat{\boldsymbol{\alpha}}_k\times\right)\right] = \left[\mathbf{I}_{3\times 3} - \left(\delta\hat{\boldsymbol{\alpha}}_k\times\right)\right] \quad (51)$$

The estimate Lie matrix $\hat{\mathbf{X}}_k$ is given by

$$\hat{\mathbf{X}}_k = \left[ \mathbf{A}(\hat{\mathbf{q}}_{k|k-1}) \mathbf{A}(\boldsymbol{\delta}\hat{\mathbf{q}}_k), \hat{\boldsymbol{\beta}}_{k|k-1} + \mathbf{A}(\hat{\mathbf{q}}_{k|k-1}) \mathbf{d}\hat{\boldsymbol{\beta}}_k \right] \tag{52}$$

It is shown that the $\mathbb{SE}(3)-\text{EKF}$ with reference attitude error bears much resemblance to the RIEKF proposed in [19]. The RIEKF is derived via the invariant Kalman filter theory which has also made use of many Lie group and Lie algebra theories. The nonlinear error-state is elaborately developed according to the invariance theory in RIEKF. In contrast, the error-state used here is derived directly from the definition of transformation matrix error.

Similar with the relationship between the GEKF and MEKF with body frame attitude error, the $\mathbb{SE}(3)-\text{EKF}$ with reference attitude error can also be viewed as the geometric algorithm of the traditional MEKF with reference frame attitude error. It can be seen from the nonlinear gyroscope bias error definition $\mathbf{d}\boldsymbol{\beta} = \mathbf{A}(\hat{\mathbf{q}})^T \Delta\boldsymbol{\beta}$ that, the linear errors in vectorspace are transformed into the reference frame using the estimated attitude. This is consistent with the attitude error definition, that is to say, all the involved errors are expressed in the same estimated reference frame.

## 5. Conclusions

In this note, the frame consistent error used in geometric extended Kalman filter has been re-derived using $\mathbb{SE}(3)$ formulation of the attitude matrix and gyroscope bias. In this respect, the geometric extended Kalman filter can be viewed as a type of $\mathbb{SE}(3)$ based extended Kalman filter for attitude estimation. Moreover, this note has also revealed an interesting link between the $\mathbb{SE}(3)$ based extended Kalman filter with reference attitude error and the right-invariant extended Kalman filter. In the further, the geometric extended Kalman filter will be further investigated making use of the invariance theory based on the $\mathbb{SE}(3)$ perspective. Meanwhile, the $\mathbb{SE}(3)$ perspective on geometric extended Kalman filter will also be extended to applications in inertial navigation.

## Appendix: Derivation of the State-Space Model with Reference Frame Attitude Error for MEKF

The multiplicative error quaternion in the reference frame is given by

$$\boldsymbol{\delta q} = \hat{\mathbf{q}}^{-1} \otimes \mathbf{q} \tag{A1}$$

Taking the time derivative of (A1) gives

$$\boldsymbol{\delta \dot{q}} = \dot{\hat{\mathbf{q}}}^{-1} \otimes \mathbf{q} + \hat{\mathbf{q}}^{-1} \otimes \dot{\mathbf{q}} \tag{A2}$$

where

$$\dot{\mathbf{q}} = \frac{1}{2}\Omega(\boldsymbol{\omega})\mathbf{q} = \frac{1}{2}\begin{bmatrix}\boldsymbol{\omega}\\0\end{bmatrix} \otimes \mathbf{q} \tag{A3}$$

The function of $\Omega(\boldsymbol{\omega})$ is explicitly defined in (A3). According to [13], the time derivative of inverse quaternion is given by

$$\dot{\hat{\mathbf{q}}}^{-1} = -\frac{1}{2}\hat{\mathbf{q}}^{-1} \otimes \begin{bmatrix}\hat{\boldsymbol{\omega}}\\0\end{bmatrix} \tag{A4}$$

Substituting (A3) and (A4) into (A2), and using the definition of the error quaternion in (A1) gives

$$\begin{aligned}\boldsymbol{\delta \dot{q}} &= -\frac{1}{2}\hat{\mathbf{q}}^{-1} \otimes \begin{bmatrix}\hat{\boldsymbol{\omega}}\\0\end{bmatrix} \otimes \mathbf{q} + \frac{1}{2}\hat{\mathbf{q}}^{-1} \otimes \begin{bmatrix}\boldsymbol{\omega}\\0\end{bmatrix} \otimes \mathbf{q}\\ &= \frac{1}{2}\hat{\mathbf{q}}^{-1} \otimes \begin{bmatrix}\boldsymbol{\delta \omega}\\0\end{bmatrix} \otimes \mathbf{q} = \frac{1}{2}\hat{\mathbf{q}}^{-1} \otimes \mathbf{q} \otimes \mathbf{q}^{*} \otimes \begin{bmatrix}\boldsymbol{\delta \omega}\\0\end{bmatrix} \otimes \mathbf{q}\\ &= \frac{1}{2}\boldsymbol{\delta q} \otimes \begin{bmatrix}\mathbf{A}(\mathbf{q})^{T}\boldsymbol{\delta \omega}\\0\end{bmatrix}\end{aligned} \tag{A5}$$

According to the definition of quaternion multiplication, (A5) can be expanded as

$$\delta \dot{\mathbf{q}} = \frac{1}{2} \begin{bmatrix} \left(\mathbf{A}(\mathbf{q})^T \delta \boldsymbol{\omega}\right) \times & \mathbf{A}(\mathbf{q})^T \delta \boldsymbol{\omega} \\ -\left(\mathbf{A}(\mathbf{q})^T \delta \boldsymbol{\omega}\right)^T & 0 \end{bmatrix} \delta \mathbf{q} \tag{A6}$$

With small angle approximation assumption, the error quaternion can be approximated as

$$\delta \mathbf{q} = \begin{bmatrix} \frac{1}{2} \delta \boldsymbol{\alpha} \\ 1 \end{bmatrix} \tag{A7}$$

Substituting (A7) into (A6) gives

$$\delta \dot{\boldsymbol{\alpha}} = \frac{1}{2} \left(\mathbf{A}(\mathbf{q})^T \delta \boldsymbol{\omega}\right) \times \delta \boldsymbol{\alpha} + \mathbf{A}(\mathbf{q})^T \delta \boldsymbol{\omega} \approx \mathbf{A}(\hat{\mathbf{q}})^T \delta \boldsymbol{\omega} \tag{A8}$$

where

$$\begin{aligned} \delta \boldsymbol{\omega} &= \boldsymbol{\omega} - \hat{\boldsymbol{\omega}} \\ &= (\tilde{\boldsymbol{\omega}} - \boldsymbol{\beta} - \boldsymbol{\eta}_v) - (\tilde{\boldsymbol{\omega}} - \hat{\boldsymbol{\beta}}) \\ &= -(\Delta \boldsymbol{\beta} + \boldsymbol{\eta}_v) \end{aligned} \tag{A9}$$

Substituting (A9) into (A8) gives

$$\delta \dot{\boldsymbol{\alpha}} = -\mathbf{A}(\hat{\mathbf{q}})^T (\Delta \boldsymbol{\beta} + \boldsymbol{\eta}_v) \tag{A10}$$

Denote the error state as $\Delta \mathbf{x}(t) = \begin{bmatrix} \delta \boldsymbol{\alpha}^T & \Delta \boldsymbol{\beta}^T \end{bmatrix}^T$ and the process noise as $\mathbf{w}(t) = \begin{bmatrix} \boldsymbol{\eta}_v(t)^T & \boldsymbol{\eta}_u(t)^T \end{bmatrix}^T$, the corresponding error model is given by

$$\Delta \dot{\mathbf{x}}(t) = \mathbf{F}(\hat{\mathbf{x}}(t), t) \Delta \mathbf{x}(t) + \mathbf{G}(t) \mathbf{w}(t) \tag{A11}$$

where

$$\mathbf{F}(\hat{\mathbf{x}}(t), t) = \begin{bmatrix} \mathbf{0}_{3\times 3} & -\mathbf{A}(\hat{\mathbf{q}})^T \\ \mathbf{0}_{3\times 3} & \mathbf{0}_{3\times 3} \end{bmatrix} \tag{A12}$$

$$\mathbf{G}(t) = \begin{bmatrix} -\mathbf{A}(\hat{\mathbf{q}})^T & \mathbf{0}_{3\times 3} \\ \mathbf{0}_{3\times 3} & \mathbf{I}_{3\times 3} \end{bmatrix} \tag{A13}$$

The next step involves the determination of the measurement transition matrix used in

MEKF. For a single sensor the true and estimated body vectors are given by

$$\mathbf{b} = \mathbf{A}(\mathbf{q})\mathbf{r} \tag{A14}$$

$$\hat{\mathbf{b}}^- = \mathbf{A}(\hat{\mathbf{q}}^-)\mathbf{r} \tag{A15}$$

According to (A1), $\mathbf{A}(\mathbf{q})$ can be obtained by

$$\mathbf{A}(\mathbf{q}) = \mathbf{A}(\hat{\mathbf{q}}^-)\mathbf{A}(\delta\mathbf{q}) \tag{A16}$$

where $\mathbf{A}(\delta\mathbf{q})$ can be approximated as

$$\mathbf{A}(\delta\mathbf{q}) = \mathbf{I}_{3\times3} - (\delta\boldsymbol{\alpha}\times) \tag{A17}$$

Substituting (A16) and (A17) into (A14) and (A15) yields

$$\begin{aligned}\Delta\mathbf{b} &= \mathbf{A}(\hat{\mathbf{q}}^-)\mathbf{A}(\delta\mathbf{q})\mathbf{r} - \mathbf{A}(\hat{\mathbf{q}}^-)\mathbf{r} \\ &= \mathbf{A}(\hat{\mathbf{q}}^-)(\mathbf{I}_{3\times3} - (\delta\boldsymbol{\alpha}\times))\mathbf{r} - \mathbf{A}(\hat{\mathbf{q}}^-)\mathbf{r} \\ &= -\mathbf{A}(\hat{\mathbf{q}}^-)(\delta\boldsymbol{\alpha}\times)\mathbf{r} = \mathbf{A}(\hat{\mathbf{q}}^-)(\mathbf{r}\times)\delta\boldsymbol{\alpha} \\ &= \mathbf{A}(\hat{\mathbf{q}}^-)(\mathbf{r}\times)\mathbf{A}(\hat{\mathbf{q}}^-)^T\mathbf{A}(\hat{\mathbf{q}}^-)\delta\boldsymbol{\alpha} \\ &= \left[(\mathbf{A}(\hat{\mathbf{q}}^-)\mathbf{r})\times\right]\mathbf{A}(\hat{\mathbf{q}}^-)\delta\boldsymbol{\alpha}\end{aligned} \tag{A18}$$

Given the error state definition, the measurement transition matrix for a single sensor can be given by

$$\mathbf{H}_k = \left[\left[(\mathbf{A}(\hat{\mathbf{q}}^-)\mathbf{r})\times\right]\mathbf{A}(\hat{\mathbf{q}}^-) \quad \mathbf{0}_{3\times3}\right] \tag{A19}$$

This matrix can be easily extended to incorporate multiple vector observations.

## Acknowledgments

This work was supported in part by the National Natural Science Foundation of China (61873275).